\documentstyle[preprint,aps,psfig]{revtex}
\tightenlines
\def\be{\begin{equation}}
\def\ee{\end{equation}}
\def\bea{\begin{eqnarray}}
\def\eea{\end{eqnarray}}
\def\ba{\begin{array}}
\def\ea{\end{array}}
\def\bc{\begin{center}}
\def\ec{\end{center}}

\def\lag{\langle}
\def\rag{\rangle}
\def\ri{{\bf r}_i}
\def\Ri{{\bf R}_i}

\def\R{{\bf R}}
\def\DE{\Delta E}

\begin{document}

\title{Dissipative Dynamics and the Statistics of Energy 
States of a Hookean Model for Protein Folding}
\author{Erkan T\"uzel$^1$,  Ay{\c s}e Erzan$^{1,2}$}
\address{$^1$  Department of Physics, Faculty of  Sciences
and Letters\\
Istanbul Technical University, Maslak 80626, Istanbul, Turkey\\
$^2$  G\"ursey Institute, P. O. Box 6, \c
Cengelk\"oy 81220, Istanbul, Turkey}
\date{\today}
\maketitle
\begin{abstract}
A generic model of a random polypeptide chain, with
discrete torsional  degrees of freedom and Hookean springs
connecting pairs of  hydrophobic residues, reproduces the energy
probability  distribution of real proteins over a very large
range of energies.   We show that this system with harmonic
interactions,  under  dissipative dynamics driven by  random
noise, leads to a  distribution of energy states obeying a
modified one-dimensional Ornstein-Uhlenbeck process and giving
rise to the
so called Wigner distribution. A tunably fine- or coarse-grained 
sampling of the energy landscape yields a family of 
distributions for the  energies and energy spacings.

PACS 5.65+b,5.70Ln,87.17.Aa
\end{abstract}
\newpage
\section{Introduction}
We present a generic  model hamiltonian for polypeptide chains 
which we believe captures the essential mechanism driving the
folding process, namely  hydrophobic 
interactions
~\cite{ben-Avraham,Dill,Tirion,Bahar,Erman1,Erman2,Tuzel}, 
and is able to reproduce the distribution of energy
states of real proteins~\cite{ben-Avraham}. 

We take the view here that the protein in its native state must 
essentially correspond to a self-organized system, i.e., the
``native state'' should be concieved of as the attractor of a
dynamics. This typically corresponds not to a unique conformation
but to a set of conformations to which the trajectory of the
phase point representing the molecule is confined after
asymptotically long times.  

Our model involves N coupled, discrete, over--damped torsional
degrees of freedom  coupled by Hookean forces and driven by
random noise. As a numerical realization of this dissipative
system,  we explore the phase space under a dynamics based on
relaxing pairs of rotational degrees of freedom, namely the
dihedral angles, sampled with a probability which is a function
of the conjugate torques,
\be 
P(i)= { \vert \tau_i\vert ^\eta\over \sum_i \vert \tau_i\vert 
^\eta}\;\;. \label{prob} 
\ee  
\noindent
The energy landscape is effectively coarse- or fine--grained by
tuning the parameter $\eta$. For $\eta=0$, the dynamics is
identical to very high temperature Monte Carlo simulations.

We find that the energy probability distribution obtained from
our  simulations 
may be very well represented by a Wigner
distribution~\cite{Wigner,Brody,Porter} which, for a random
quantity $S$, 
is given by
\be 
P(S)\sim S\; \exp(- {\pi \over 4} S^2)\;\; \label{Wigner}
\ee 
while the coarse
grained energy level distributions are comparable with the 
$n$'th ($n=1,2,3\ldots$) neighbor energy level spacing statistics
encountered~\cite{Porter1} in the study of large nuclei. The
energy histograms can also be very well fitted with an ``inverse 
Gaussian" (IG) distribution.  

We are able to show that for harmonic potentials, quite
independently  of the nature of the sequence of hydrophobic and
polar residues, or the dimensionality of the space, the
energy of the system obeys a modified Ornstein-Uhlenbeck (OU)
process~\cite{Feller}.  
The stationary state distribution for this process with
reflecting boundary conditions introduced due to constraints, may
then be related to the energy distribution.
As a  bonus,
we are also able to understand the
distribution of relaxation times found for global optimization
problems~\cite{Li} by Li and coworkers. 

We have already reported in a separate publication \cite{Tuzel}
that  under Metropolis Monte Carlo dynamics, with random initial
conditions,  the model exhibits  power law relaxation for the
initial stages of decay,  and at the later stages the relaxation
obeys a stretched exponential with the exponent $\beta\simeq 
1/4$.  This  Kohlrausch-Williams-Watts type relaxation behaviour
is observed experimentally for real proteins
~\cite{Wolynes1,Angell,Bahar1,Colmenero}. At zero temperature
the probability distribution function of the energy steps
encountered along a relaxation path in phase space also obeys a 
stretched exponential form, with another exponent $\alpha\simeq
0.39$. In~\cite{Tuzel} we show that $\beta=\alpha / (\alpha +
1)$,
which yields a value for $\beta$ in very good agreement with
our simulation results. 

The paper is organized as follows.  In section 2 we define our
model, in
section 3 we present our simulation results, in section 4 we show
that the
energy obeys an OU rocess. In section 5 we discuss the
relationship
between this model and other complex systems and outline work in
progress. 
  
\section{The Model} 
We consider a model~\cite{Tuzel} consisting of N
residues, treated as point vertices, interacting via Hookean
potentials.
We have been motivated by the model proposed by Halilo{\u g}lu,
Bahar,
Erman~\cite{Bahar} where all interactions between the different
residues
are governed by confining square-law potentials
~\cite{Bahar,Erman1,Erman2}.  In our model, however, the covalent
bonds
between residues are treated as fixed rods of equal length (see
Fig.1). 
The residues located at the vertices may be polar $P$ or
hydrophobic $H$.
All the hydrophobic vertices are to be connected to each other
with
springs of equal stiffness. This feature mimicks the 
effective pressure that is
exerted on the hydrophobic residues by the ambient water
molecules, and
results in their being driven to the relatively less exposed
center of the
molecule in the low lying energy states, whereas the polar
residues are
closer to the surface. It is important to note that we
treat all $H-H$ pairs on an equal footing, i.e., there is no
``teleological" information that is fed into the system by
connecting only
those $H-H$ pairs which are close to each other in the native
configuration for a particular sequence.  It is known that real
proteins
are distinguished by $H-P$ sequences that lead to unique ground
states
while a randomly chosen $H-P$ sequence will typically give rise
to a
highly degenerate ground state. In the absence of detailed
knowledge regarding
the rules singling out the realistic $H-P$ sequences we
considered a
generic $H-P$ sequence obtained by choosing fifty percent of the
residues
to be hydrophobic and distributing them randomly along the chain.
We have
checked that our results were quite robust with respect to
changing the
sequence of hydrophobic or hydrophilic residues, or even taking
all of
them to be hydrophobic. 

The energy of the molecule is
\be
E= {K\over 2} \sum_{i,j} c_{i,j} \vert {\bf r}_i-{\bf r}_j\vert^2
= K \sum_{i,j} {\bf r}_i^{\dagger} V_{ij} {\bf r}_j
\label{energy}
\ee
If we define $Q_i=1$ for the $i$' th vertex being occupied by a
hydrophobic  residue, and $Q_i=0$ otherwise, we may write
$c_{i,j}=Q_i Q_j$ and  
\begin{eqnarray}
V_{ij}=[(&N&_H-1) c_{i,i} -c_{i,j-1}-c_{i,j+1}]\delta_{i,j} \cr
& -& (1-\delta_{i,j})(1-\delta_{i,j-1}-\delta_{i,j+1}) c_{i,j}
\;\;\;.\label{Vij}
\end{eqnarray}
\noindent

We take the bond angles $\alpha_i, i=1\ldots, N-1$, to  have the
alternating values of  $(-1)^i \alpha,$ with $\alpha=68^\circ$.
The dihedral angles $\phi_i$ can take on the values of 0 and $\pm
2\pi/3 $.
The state (conformation) of the system is uniquely specified once
the numbers $\{\phi_i\}$ are given. 
The constraints placed on the conformations due to the rigid
chemical
bond lengths and by restricting the chemical and dihedral angles
to
discrete values prevent the molecule from trivially collapsing to
a point.
The residues effectively reside on the vertices of a tetrahedral
lattice.
The position vectors ${\bf r}_i$ of each of the vertices in the
chain can
be expressed in terms of a sum over the directors $\Ri$ of unit
length
representing the chemical bonds, which may be obtained from ${\bf
R}_1$ by
successive rotations ${\bf M}_k(\alpha_k)$ and ${\bf
T}_k(\phi_k)$ through
the bond and the dihedral angles~\cite{Flory}, viz., 
\be 
\ri=
\sum_{j=1}^{i-1} \prod_{k=j}^{2} {\bf T}_k(\phi_k) {\bf
M}_k(\alpha_k)
\R_1\;\;\;.  
\ee where we may choose $\R_1$ to lie along any of the
Cartesian directions in our laboratory frame without loss of
generality. 
We obtain the torques that act at each of
the vertices $i$ by substituting this in equation (\ref{energy})
and
taking the partial derivative with respect to $\phi_i$, viz., 
\be
\tau_i=-\partial E/ \partial \phi_i \;\;.  
\ee 

The system is assumed to evolve within a viscous environment,
subject to random kicks from the ambient molecules.  We may write
the Langevin equation,
\be
{d {\bf r}_i(t) \over d t}= {1 \over \zeta_r}
{\bf F}_i+ {\bf \xi}_r(i,t)\label{Langevinr}\ee
where $\zeta_r$ is a friction coefficient and {\boldmath
$\xi$}$_r(i,t)$
is a Gaussian distributed noise term, delta correlated in $i$ and
in time.
Equivalently, in terms of the state vector
{\boldmath ${\phi}$} $=(\phi_1,\ldots,\phi_N)$, we have  the
Langevin
equation 
\be
{d { \phi_i(t)} \over d t}={1\over \zeta_\tau} 
\tau_i + {\xi}_\tau (i,t)\label{Langevin}\ee
where the torque ${\tau_i}$ is a  function of all the angles
$\{${\boldmath $\phi$}$\}$, $\zeta_\tau$ is the appropriate
friction
coefficient
and ${\xi_\tau}$ is again a Gaussian random force delta
correlated in
space and time.  Viewed in this way the dynamics is similar to a
pinned interface or a charge density wave
system~\cite{Erzan,Veermans,Parisi}  in $1+1$ dimensions. 

For the discrete, sequential numerical simulation of the
evolution of this system, we postulate the following set of
rules:
\begin{enumerate}
\item
Form the self-similar probability distribution in equation
(\ref{prob}), 
$ P(i)=  \vert \tau_i\vert ^\eta/ \sum_i \vert \tau_i\vert 
^\eta$ 
\item 
Choose a pair of  vertices ($i,i^\prime$) according to  this
probability distribution  over $\{\tau_i >0\}$ and
$\{\tau_i<0\}$, 
\item 
Set $\phi_i(t+1)=\phi_i(t) + \;{\rm sign}(\tau_i)(2\pi/3)\;.$ 
\end{enumerate}

Here $\eta$ is a tunable parameter defining the dynamics. For 
large positive values of $\eta$, those angles $\phi_i$ with  the 
maximal conjugate torques are incremented; for negative values 
of $\eta$ the small values of the torque are preferred.  For
$\eta = 0$ the angles to be incremented are picked randomly. If
one choses
$\eta$ to be very large, then we find that there is a large
probability that the most recently updated $\phi_i$ still carries
a very large torque, resulting in a jamming of the dynamics.
Incrementing the dihedral angles with the large conjugate torques
resulted not in the relaxation of these torques but in pumping
energy into the system, as when pushing a swing at the top of its
arc.  After applying the search strategy based on changing the
torques according to a distribution, we found that updating the
maximal torques ($\eta >0$) drives the system to a state with
relatively high energies, whereas a random search ($\eta =0$) or
preferentially choosing the minimal torques ($\eta <0$) gives
rise to  more successful strategies for reaching low lying energy
states.  Thus it can be said that $\eta$ here plays the role of
a coarse-- or fine--graining parameter in the exploration of the
energy landscape. 

\section{Distribution of energy states and level spacings}

The distribution of the energies of the discrete configurational
states explored by the chain of $N= 48$ residues shown in Fig.1, 
as it evolves under the above dynamics, is shown in Figs.2-5, 
for both positive and negative $\eta$. After the first 5000
steps were discarded, the statistics were taken over 5000 steps
of the trajectory.  It can be seen that the shape of the curve
does not essentially change, while for positive
$\eta$ the peak shifts to successively higher values of the
energy, and the distribution is distorted towards a Gaussian,
indicating that
the states explored are less correlated. 
These figures should be compared with those reported by 
ben--Avraham~\cite{ben-Avraham} for the density of vibrational
states
and by Mach et al.~\cite{Mach} for the ultraviolet
absorption spectra, and also with the energy histograms obtained
by Socci and 
Onuchic~\cite{Onuchic} for a Monte Carlo simulation on a lattice
model of 
proteinlike heteropolymer. Our computation seems to be very
successful in 
producing realistic distributions of energy states over the whole
range
of relevant energies.

We have been able to fit the simulation results very successfully
with a distribution of the Wigner form (Figs.
2,3)
\be 
f_{\rm W}(E)=a (E-E_0) e^{-b (E-E_0)^2} \;\;, \label{Wigner1}
\ee 
\noindent
for $\eta=-6$ to $\eta=3$,  and the parameters for the fit are
given in Table Ia. Here $E_0$ corresponds to the 
offset due to the lowest energy state attained for the different
$\eta$, and it can be seen that it shifts the distribution to
higher values of the energy for higher values of $\eta$. The
distributions become Gaussian for $\eta=6$ and $\eta=8$; the
results of the Gaussian fits are presented in Table Ib.
 
It should be mentioned that the same energy distributions may be 
fitted equally well or better by  the ``inverse
Gaussian''~\cite{Li}, where
the  probability density is given by (see Figs. 4,5), 
\be 
f_{\rm IG}(E)=\sqrt{{A \over 2\pi E^3}}\;
\exp\left[-{A(E-B)^2\over 2 B^2 E}\right]\;\;. \label{IG1}
\ee 
It will be noted that this has the same functional form as the
distribution of first passage times over a distance $d$ for an
Ornstein Uhlenbeck process~\cite{Feller} with diffusion 
coefficient $D$ and initial  drift velocity  $v$, in the regime
of small times, if one makes the further  identifications
$A=d^2/(2D)$ and $B=d/v$.  We postpone until section 4 a
discussion of this result.  The parameters for the  fits to the
parameters $A$ and $B$ are given in Table II.  The estimated
errors for each fit are also given in the table. We find that
both the ``diffusion constant (mobility)'' and the ``drift
velocity'' of the phase point  along its trajectory in phase
space depend on $\eta$, being maximum for $\eta=0$  and
decreasing for positive values of $\eta$. For $\eta < 0$ they
essentially stay  the same.

We have also considered the statistics of energy differences 
between successive energy states visited along a trajectory 
obeying the above dynamics. This does not necessarily mean that 
the energy differences considered here are nearest neighbors on 
the energy spectrum; rather these statistics may be considered 
a classical analogue of an absorption (or emission) spectrum.  
We found that the distributions were symmetric for $\Delta E$ 
negative or positive, and that they obeyed a stretched 
exponential distribution (for positive $\Delta E$),  
\be 
P(\Delta E) \sim \exp [-(\Delta E)^c)] \;\;.
\ee 
The distribution of energy steps for different $\eta$ are given
in Fig.6a. The plots of $\ln(-\ln(P(\Delta E)))$ versus
$\ln(\Delta E)$ for $\eta=0$ and $\eta=8$ are shown in Fig.6b.
The fits for other $\eta$ values are equally good.  The values
of the exponent $c$ are shown in Table III, where it can be seen
that $c$ starts from small values  for $\eta=-8$ and seems to
tend to 1 as $\eta$ becomes large and positive, again exhibiting
a decorrelation effect as the energy landscape is probed with
larger and larger $\eta$. The correlation coefficients showing
the goodness of fit for each $\eta$ are also given in Table III.

Finally, it is interesting to make a comparison between the
energy states explored under the ``$\eta$'' --dynamics and
Metropolis Monte Carlo. Since
\be 
P(E)=n(E) \exp (-\gamma E)\;\;\;,\ee
$n(E)$ should become identical to $P(E)$ in the limit of
$\gamma \to 0 $.
In Fig.7 we show our results for $n(E)$ with $\gamma = 10^{-5}$, 
and $P(E)$ for $\eta=0$.

\section{Ornstein-Uhlenbeck Process and the Wigner Distribution}

We would now like to show that the energy obeys a
stochastic process which can be modelled by
Fokker-Planck equations with the Wigner~(\ref{Wigner1}) or the
inverse
Gaussian~(\ref{IG1}) forms as stationary solutions.

We remind the reader that an OU process
describes the diffusive motion of a particle subject to a drift
velocity proportional to the distance from the
origin~\cite{Feller}.  It can easily be seen that such a process
for a single particle in one dimension would be described by the
Langevin equation,
\be
{d x \over d t}= -{1\over \zeta} gx +
\xi(t)\label{Langevin1}\ee
with a Hookean force $F(x)=-gx$ and a delta correlated random
force $\xi(t)$, $\lag (\xi(t))^2\rag = \sigma^2$.  

Since there is no explicit time dependence of $E$, we have 
\be
{d E\over dt}=\sum_i {\partial E \over \partial {\bf
r}_i}\cdot{\partial
{\bf r}_i \over \partial t}\;\;.\label{OU1}\ee
Substituting from (\ref{Langevinr}) we get,
\be
{d E\over dt}=-{1\over \zeta_r} \sum_i \left({\partial E\over
\partial {\bf r}_i}\right)^2 + \sum_i {\partial E
\over \partial {\bf r}_i}\cdot
{\bf \xi}_i(t)\;\;.\label{OU2}\ee
From (\ref{energy}) we may compute that 
\be 
\sum_i \left({\partial E\over \partial {\bf r}_i}\right)^2=
{NE\over
\zeta_r}+
\sum_{i,j,k \atop{i\ne j}} c_{ik}c_{jk} ({\bf r}_i-{\bf
r}_k)\cdot ({\bf r}_j-{\bf r}_k)\;\;.\ee
We see that the second term is like an average of the products 
$({\bf r}_i-{\bf r}_k)\cdot ({\bf r}_j-{\bf r}_k)$ over
$(i,j)$ pairs $(i\ne j)$,
and for a reasonably isotropic configuration, it vanishes. To the
same
approximation, we may assume that  the second term in
Eq.(\ref{OU2}) is itself equal to a Gaussian stochastic noise,
i.e.,  set $\xi_E(t) =K\sum_{ij}c_{ij} ({\bf r}_i-{\bf
r}_j)\cdot {\bf \xi}_i(t)\;\;.$
This yields the required result, namely, 
\be
{d E\over dt}=-{NE\over \zeta_r} + \xi_E(t)\;\;\;.\label{OU3}
\ee

This stochastic equation is equivalent~\cite{Risken} to the 
Fokker-Planck equation
\be
{\partial P(E,t)\over \partial t} =-{\partial \over \partial E}
\left[-{\partial \tilde{\Phi}(E)\over \partial E} P(E,t) -D
{\partial
P(E,t)\over \partial E}\right] \;\;, \label{FP}\ee
for the probability distribution of $E$, 
where $D=2\langle \xi_E^2\rangle$
and $\tilde{\Phi}(E)=\int_0^E (N/\zeta_r)\,x\,dx=b{E^2}/2$, with
$b=N/\zeta_r$.  
The constraints we have placed on our
configurational degrees of freedom (see Eq.(\ref{Vij})ff.)
require that 
there be some minimum value of the energy where the probability
current
vanishes, implying reflecting boundary conditions there, as well
as at
some $E_{\rm max}$, which we may take to $\infty$ for all
practical
purposes. To mimick these constraints we introduce an infinitely
high
potential barrier at $E_0$, while at the same time shifting the
point of
equilibrium of the Hookean ``force'' to this point.  A convenient
choice
for a singular potential to add to $\tilde{\Phi}$, is
$-\ln(E-E_0)$. 
These reflecting boundary conditions at $E_0$ and at $\infty$ 
then lead to a stationary solution $P(E)$,
\be
P(E)=\,a\,e^{-\Phi(E)}\;\;\;,\label{PEst}\ee
where $\Phi(E)=\tilde{\Phi}-\ln(E-E_0)$, or,
\be
\Phi(E)={1\over 2} b (E-E_0)^2-\ln(E-E_0)\;\;\,\label{quadln}\ee
and $a$ is a normalization constant. Substituting (\ref{quadln})
in
(\ref{PEst}) leads to the Wigner formula~(\ref{Wigner1}).

A stationary distribution of the inverse Gaussian form may be
obtained
if we modify the quadratic potential $\tilde{\Phi}$ in a
different way
to model the constraints in the system, viz., 
\be
\Phi(E)= {A \over 2 B^2 E} (E-B)^2 +{3\over 2} \ln E \;\;\;.\ee
This also leads to reflecting boundary conditions, at $E=0$ and
$E\to
\infty$, and a point of equilibrium at $E=B$. As the stationary
solution
we obtain the inverse Gaussian distribution~)\ref{IG1}), as can
be seen
from
direct substitution into (\ref{PEst}).

The distribution of first passage times for the attainment of the
optimum 
solution in such diverse high dimensional optimization problems
as fits
to X-ray patterns, travelling salesman problems and determination
of the 
lowest energy state for lattice models of protein configurations,
have 
been reported by Li and coworkers~\cite{Li}, to obey the
Ornstein-Uhlenbeck
form. The plots of these distributions all display a striking
similarity
to each other, and to the distribution of energy states which we
have
found in the present problem. Now we see that if an optimization 
problem has a quadratic cost function $C$ which, in terms of the
large
number of variational parameters in a reasonably isotropic phase
space,
has the same form as our energy  
Eq.(\ref{energy}), then the optimization algorithm defines a 
{\em dynamics} for $C$ which may be described by means of an OU
process 
as in Eq.(\ref{Langevin1}), with a repulsive barrier at $C_{\rm
min}$ and
at $\infty$. This
may be modelled by the same Fokker
Planck equation~(\ref{FP}), and potentials $\Phi$, as we have
discussed
above.

Recall that for  an OU process, with an initial displacement 
$x(0)=d$,
the solution
for the
distribution of first passage times $t$ through the origin is
given
by~\cite{Feller},
\be
f(t)={2yd\over \pi^{1/2} \sigma} \left({\rho \over 1-
y^2}\right)^{3/2} e^{-{\rho y^2 d^2\over \sigma^2 (1-
y^2)}}\;\;,\label{FPT}\ee
where $\rho=g/\zeta$ and $y=\exp(-\rho t)$. 
We see that (\ref{FPT}) goes over, in the
limit of large times, i.e. $y\ll 1$, to
\be
f_{\rm W} (y) = {2  d \rho^{3/2}\over \pi^{1/2} \sigma} y\,
\;e^{-{\rho d^2 y^2 \over \sigma^2}}\;\;\;.\label{Wigner2}\ee
On the other hand, for very small times, (\ref{FPT}) becomes, to
leading order,
\be 
f_{\rm IG} (t) ={2 \pi d \sigma^2 \over (2 \pi \sigma^2 t)^{3/2}}
e^{-
{(d-vt)^2\over 2\sigma^2 t}}\label{IG2}\ee
where we have defined $\rho d=v$. It should be noted that
these functions (\ref{Wigner2} and \ref{IG2}) 
have the same form, as functions of $y$ and $t$, respectively, as
the Wigner and inverse Gaussian distributions which we have found
above, and numerically are very similar to each other.

\section{Discussion}
The experimentally reported density of vibrational energy  states
and UV absorption spectra for a large number of real proteins 
fall on a universal curve~\cite{ben-Avraham,Mach}, which
we have been able to fit extremely well by distributions of the
Wigner 
or inverse Gaussian form. Our simple coarse-grained model for the
hydrophobic interactions reproduces very closely the same
distribution of energy  states. Moreover, this same curve appears
to describe the distribution of first passage times in 
rather general global optimization problems~\cite{Li}. 
We have provided an explanation for this fact 
on the basis of the Hookean model of hydrophobic
interactions.

We would like to make a few concluding remarks about about a
remaining puzzle, which is the connection between the Wigner
distribution, which arises as the distribution of eigenvalue
spacings for Gaussian orthagonal
matrices~\cite{Wigner,Brody,Porter,Mehta}, and the present
limiting form we have found for the distribution of energy
states.  The puzzle promises to be fruitful, because 
we believe that the current state of the art of understanding the
energy
landscapes of proteins is very similar to that of the study of 
the energy spectra of very large nuclei in the '50s, when it was 
realized that the problem might be very advantageously treated 
as a statistical one. For real, large nuclei, the level spacing
distribution,
properly scaled with the average density of states for different
energies,
exhibit a remarkable invariance over the entire energy
range~\cite{Brody}.
It was then  found that the energy levels and nearest-neighbor
spacings are
governed to a large extent by the statistics of eigenvalues and
eigenvalue
spacings of orthogonal matrices with a Gaussian distribution of
matrix elements~\cite{Wigner,Brody,Porter,Mehta}. The statistics
of nuclear
energy level spacings approximately obey the so called  Wigner
``surmise'' (\ref{Wigner}).
This is the distribution of eigenvalue spacings for $2 \times 2$
Gaussian 
orthogonal matrices which can be also obtained by considering the
square
root of the sum of the squares of two independent but identically
distributed Gaussian random variables of zero mean.

It is of great interest to note that the so-called Wigner
distribution
arises also in ``quantum
chaos''~\cite{Berry,Yurtsever1,Yurtsever2} and in the
energy spectra of large atomic clusters~\cite{Wales}. Thus, there
seems to be a universality to the spectral fluctuations of
confined
systems of sufficient complexity.~\cite{Brody,Bohigas}

It is very intrigueging to compare the results for $\eta\ge0$
(Fig. (2)) with  the numerically obtained $n$th neighbor 
spacing distributions
of the eigenvalues for   Gaussian orthagonal
matrices, as reported by Porter~\cite{Porter1}, where  the
identical shift of the peak and tendency to a symmetric Gaussian
distribution  is found. This we interpret as reinforcing our
observation that larger $\eta$ dynamics results in a more
coarse-grained sampling of the energy landscape. Nevertheless,
the connection to $n$'th neighbor spacing distributions of
eigenvalues of random matrices and the Ornstein-Uhlenbeck process
still remains to be understood.

\bigskip
{\bf Acknowledgements}

We wish George Stell many a happy return on the occasion of his 
65th birthday. 
It is a pleasure to thank Ersin Yurtsever for pointing out to us
the similarity between our numerical results and the Wigner
distribution. We are grateful to Burak Erman for many useful
discussions  
and for bringing Ref.~\cite{Li}
to our attention. One of us (A.E.) acknowledges partial support
by the Turkish Academy of Sciences.

\newpage
\bc
{\bf TABLE CAPTIONS}
\ec
\noindent
{\bf Table Ia} The parameters $a,b$ and $E_0$ used for fitting
the energy histograms to the Wigner distribution   
$f_{\rm W}(E) = a (E-E_0) e^{-b (E-E_0)^2} $. 

\noindent
{\bf Table Ib} The mean $\tilde{a}$ and the variance $\tilde{b}$ 
used for fitting the energy histograms to the Gaussian
distribution.  

\noindent
{\bf Table II} The parameters $A,B$ used for fitting the energy
histograms to the inverse Gaussian distribution (see
Eq.(\ref{IG1})). 
The estimated errors are
also provided. (Calculated using Levenberg-Marquart algorithm)

\noindent
{\bf Table III} The parameter $c$ used for fitting the
distribution of energy steps to a stretched exponential in the 
form  $P(\Delta E) \sim \exp [-(\Delta E)^c]$. 
The correlation coefficients ($r^2$) are also provided.  

\newpage
\bc
{\bf FIGURE CAPTIONS}
\ec
\noindent
{\bf Fig. 1} A chain of $N=48$ residues, half of which are
randomly chosen to be
hydrophobic, (darker beads)
shown in a random initial configuration.
(Generated using RasMol V2.6)

\noindent
{\bf Fig. 2} The normalized energy histograms, 
averaged over $10$ random initial states for chains of $N=48$, 
for different $\eta \ge 0$, along paths of $10^4$ steps, with the

first $5000$ steps discarded. The
fits are to the Wigner distribution for $\eta = 0, 1, 3$ and 
Gaussian distribution for $\eta=8$.

\noindent
{\bf Fig. 3} The normalized energy histograms, for chains of
$N=48$, 
for different $\eta < 0$  (see Fig.2). The fits are to the Wigner
distribution.

\noindent
{\bf Fig. 4} The normalized energy histograms along trajectories 
in phase space for the $N=48$ chain, for $\eta \ge 0$ as in
Fig.2, 
fitted with the ``inverse Gaussian'' distribution given in
Eq.~(\ref{IG1}).

\noindent
{\bf Fig. 5} Energy histograms for $\eta < 0$ as in Fig.4, fitted
with
the ``inverse Gaussian'' distribution given in Eq.~(\ref{IG1}),
for the 
$N=48$ chain.

\noindent
{\bf Fig. 6a} The distribution of energy steps along a trajectory
in phase
space according to the $\eta$- dynamics of the $N=48$ chain, for
different 
$\eta$.
The last $5000$ steps along a $10000$ steps trajectory were
considered.

\noindent
{\bf Fig. 6b} The fits, for $\eta = 0$ (left) and for $\eta=8$
(right) to
the stretched exponential form $\sim \exp(-\DE^c)$, for $c=0.58$ 
and $c=0.81$ respectively, in the large $\DE$ limit.

\noindent
{\bf Fig. 7} Energy histogram for $\eta=0$ and the density of
states,
$n(E)$, for $\gamma=10^{-5}$ obtained for $\eta$- dynamics and
the Metropolis Monte 
Carlo respectively.

\newpage
\begin{table}
{\bf Table Ia}
\bc
\begin{tabular}{|c|c|c|c|}
\hline
{\boldmath ~~~~~$\eta$~~~~~} & $\bf a \;\; (10^{-4})$ & $\bf b
\;\; (10^{-7})$ & ~~~~$\bf E_0$ ~~~~ \\ \hline\hline
-6 &       1.50 &       15.0 &        420 \\ \hline
-4 &       1.50 &       15.0 &        380 \\ \hline
-2 &       2.00 &       15.0 &        350 \\ \hline
 0 &       1.25 &        8.7 &        300 \\ \hline
 1 &       0.40 &        2.0 &        950 \\ \hline
 3 &       0.37 &        1.2 &       1300 \\ \hline
\end{tabular}  
\ec
\end{table}
\newpage
\begin{table}
{\bf Table Ib}
\bc
\begin{tabular}{|c|c|c|}
\hline
{\boldmath ~~~~~$\eta$~~~~~} & {\boldmath ~~~~~$\tilde{a}$~~~~~}
&
{\boldmath ~$\tilde{b} \;\; (10^6)$~} \\ \hline\hline
6 &    4300 &       2.2 \\ \hline
8 &    4800 &       2.7 \\ \hline
\end{tabular}  
\ec
\end{table}

\newpage
\begin{table}
{\bf Table II}
\bc
\begin{tabular}{|c|c|c|c|c|}
\hline
{\boldmath ~~~~~~~$\eta$~~~~~~~} & {\bf ~A \boldmath $(\times
10^3)$~} & {\boldmath $\Delta A(\times 10)$} & {\bf ~B
\boldmath $(\times 10^3)$~} & {\boldmath $\Delta B(\times 10)$}
\\ \hline \hline
-6 &       7.4 &        48.0 &       1.2 &         2.3 \\ \hline
-4 &       6.8 &        10.0 &       1.2 &         0.5 \\ \hline
-2 &       6.6 &         6.4 &       1.1 &         0.3 \\ \hline
 0 &       6.3 &         4.9 &       1.4 &         0.4 \\ \hline
 1 &      18.4 &         7.9 &       3.1 &         0.8 \\ \hline
 3 &      28.3 &        32.2 &       4.0 &         1.2 \\ \hline
 6 &      33.7 &        39.5 &       4.6 &         1.4 \\ \hline
 8 &      38.3 &        59.2 &       5.2 &         2.0 \\ \hline
\end{tabular}  
\ec
\end{table}

\newpage
\begin{table}
{\bf Table III}
\bc
\begin{tabular}{|c|c|c|}
\hline
\boldmath ~~~~~~~~~$\eta$~~~~~~~~~ & ~~~~~~~~~~{\bf c}~~~~~~~~~~
& $\bf r^2\;\;(corr. coef.)$ \\ \hline\hline
-8 &     0.50 &     0.89 \\ \hline
-6 &     0.49 &     0.97 \\ \hline
-4 &     0.54 &     0.98 \\ \hline
-2 &     0.54 &     0.98 \\ \hline
 0 &     0.58 &     0.97 \\ \hline
 1 &     0.74 &     0.95 \\ \hline
 2 &     0.73 &     0.96 \\ \hline
 3 &     0.81 &     0.95 \\ \hline
 4 &     0.73 &     0.96 \\ \hline
 6 &     0.85 &     0.95 \\ \hline
 8 &     0.81 &     0.95 \\ \hline
\end{tabular}  
\ec
\end{table}

\newpage

\begin{figure}
\begin{center}
\leavevmode
\psfig{figure=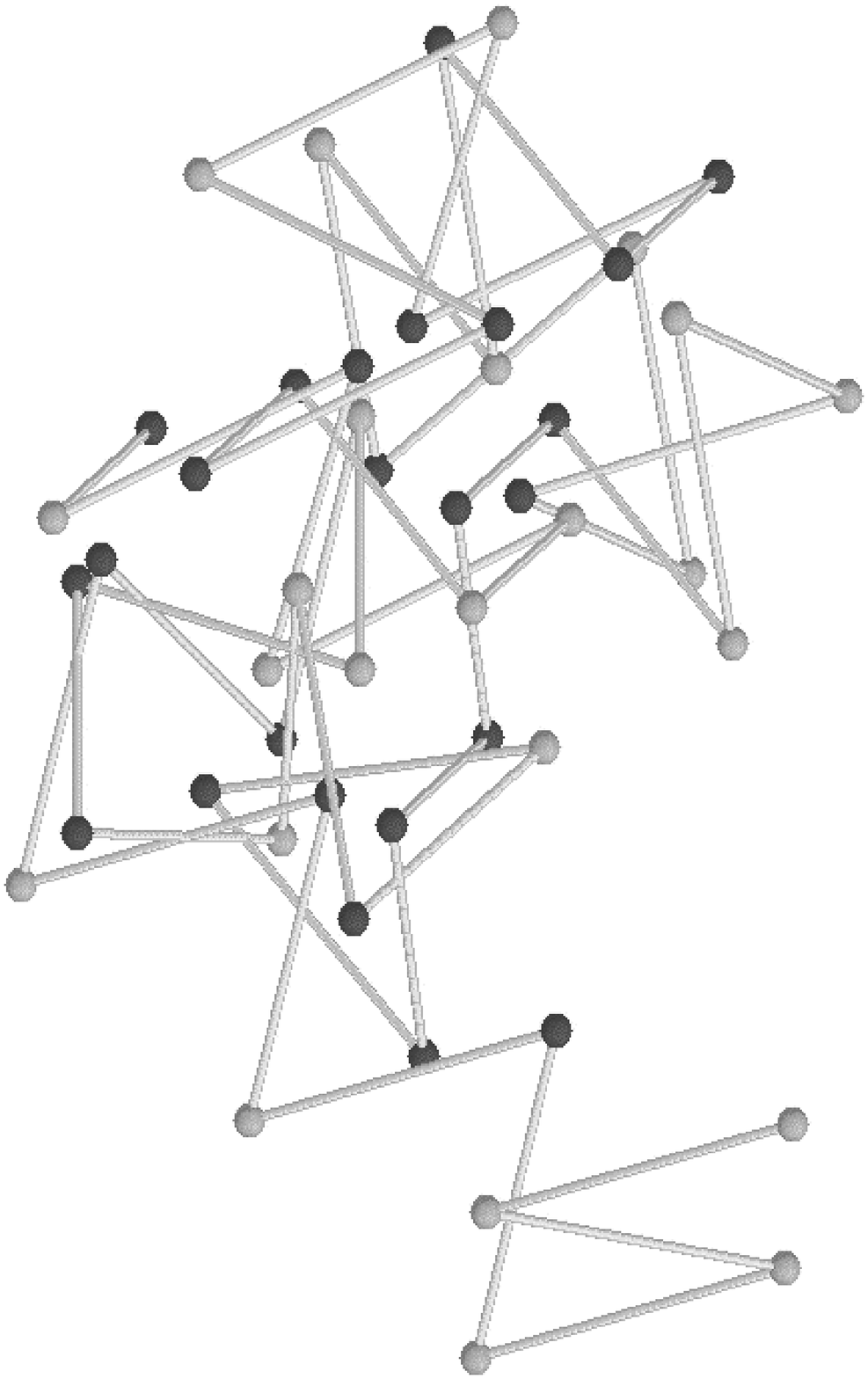,width=15cm,height=20cm,angle=0}
\end{center}
\end{figure}
\bc
{\bf Figure 1}
\ec

\newpage

\begin{figure}
\begin{center}
\leavevmode
\psfig{figure=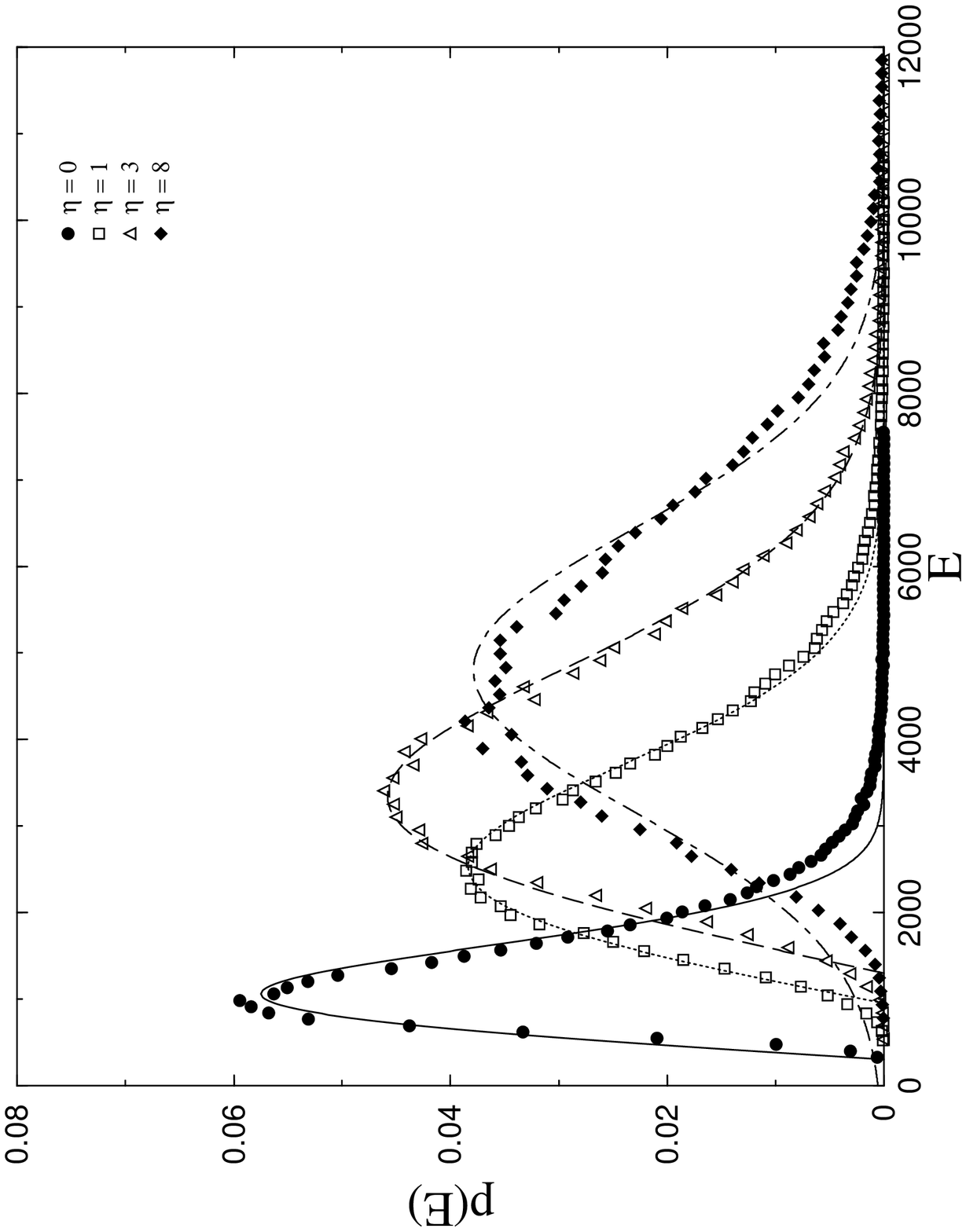,width=15cm,height=20cm,angle=0}
\end{center}
\end{figure}
\bc
{\bf Figure 2}
\ec

\newpage

\begin{figure}
\begin{center}
\leavevmode
\psfig{figure=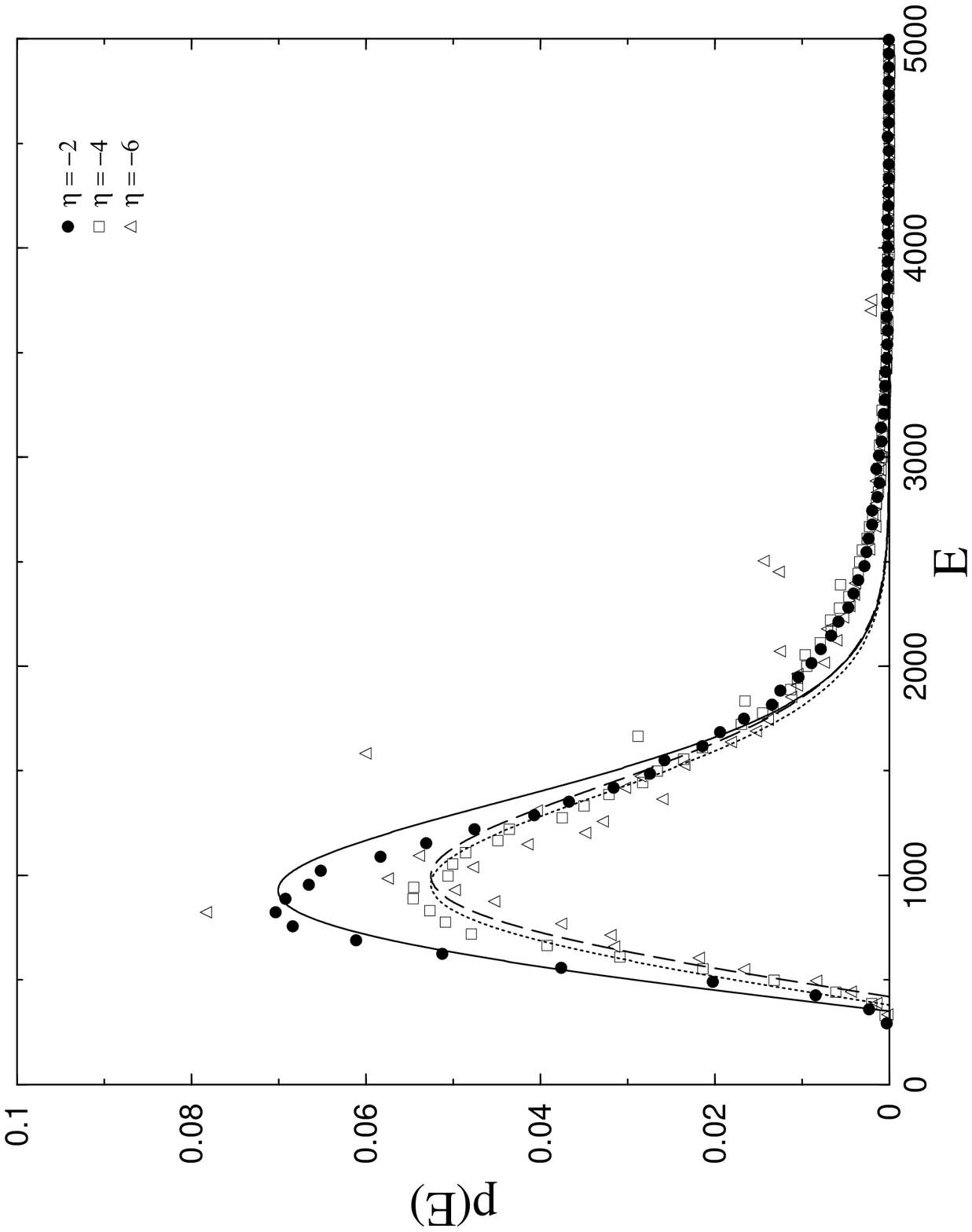,width=15cm,height=20cm,angle=0}
\end{center}
\end{figure}
\bc
{\bf Figure 3}
\ec

\newpage
\begin{figure}
\begin{center}
\leavevmode
\psfig{figure=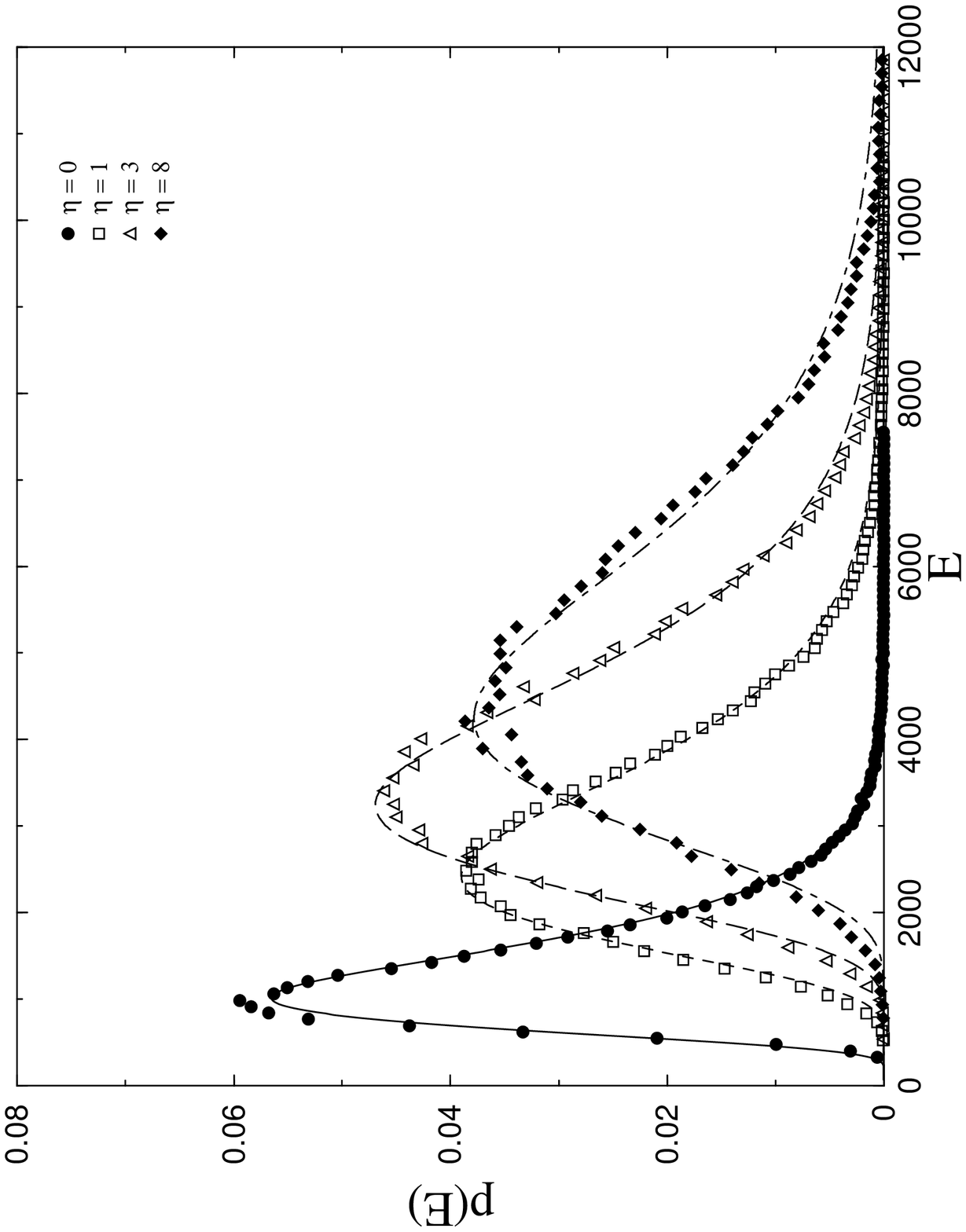,width=15cm,height=20cm,angle=0}
\end{center}
\end{figure}
\bc
{\bf Figure 4}
\ec

\newpage
\begin{figure}
\begin{center}
\leavevmode
\psfig{figure=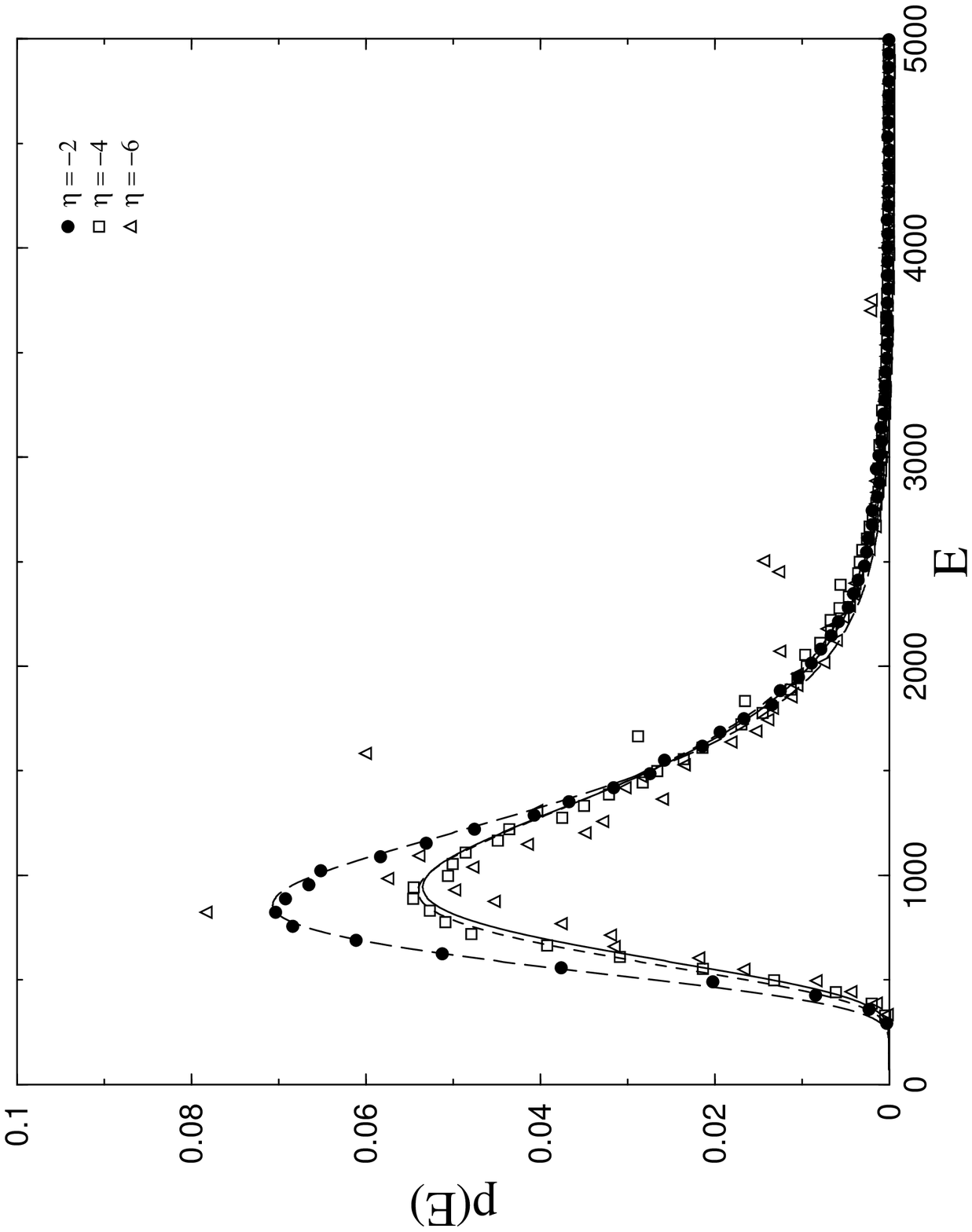,width=15cm,height=20cm,angle=0}
\end{center}
\end{figure}
\bc
{\bf Figure 5}
\ec

\newpage
\begin{figure}
\begin{center}
\leavevmode
\psfig{figure=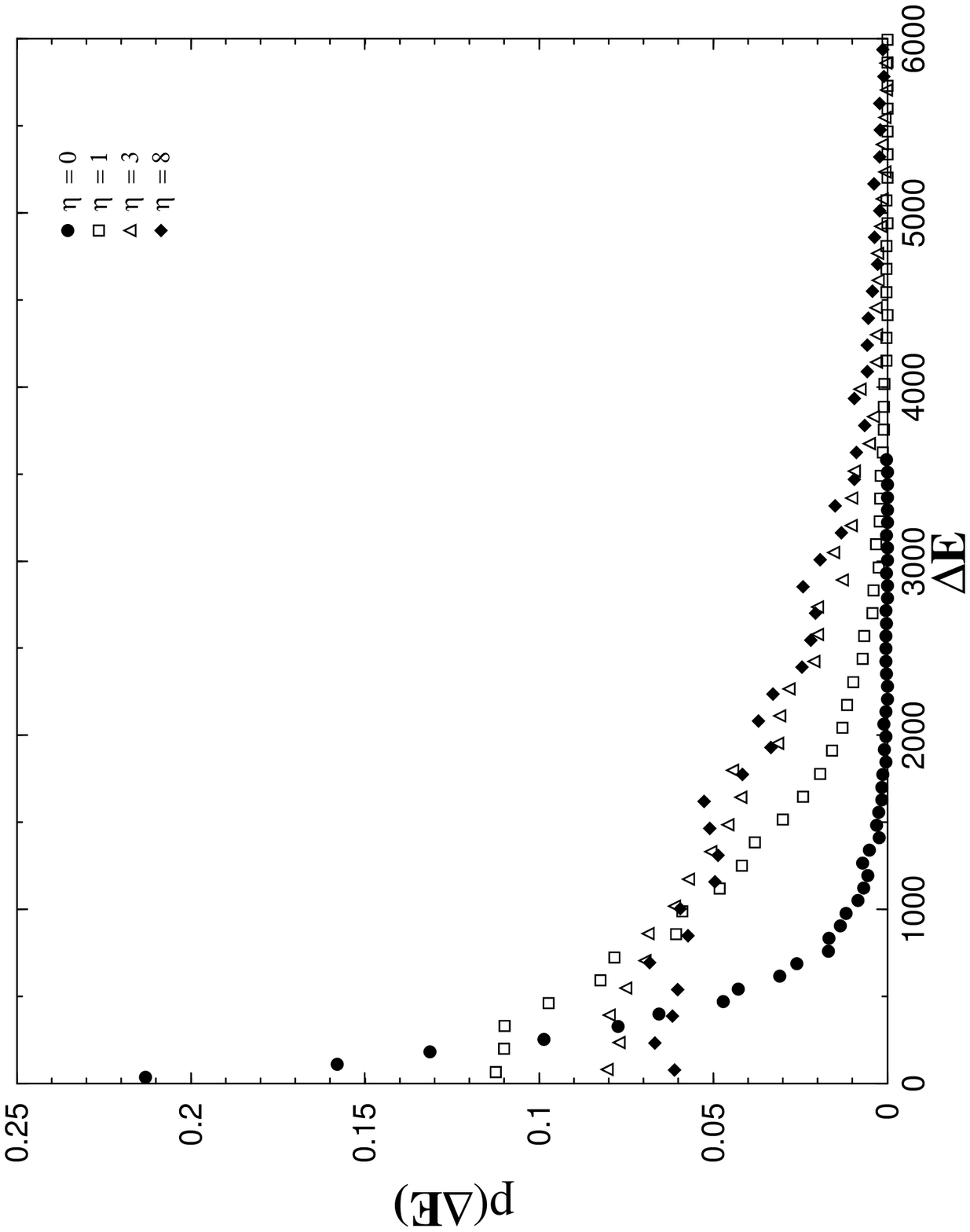,width=15cm,height=20cm,angle=0}
\end{center}
\end{figure}
\bc
{\bf Figure 6a}
\ec

\newpage
\begin{figure}
\begin{center}
\leavevmode
\psfig{figure=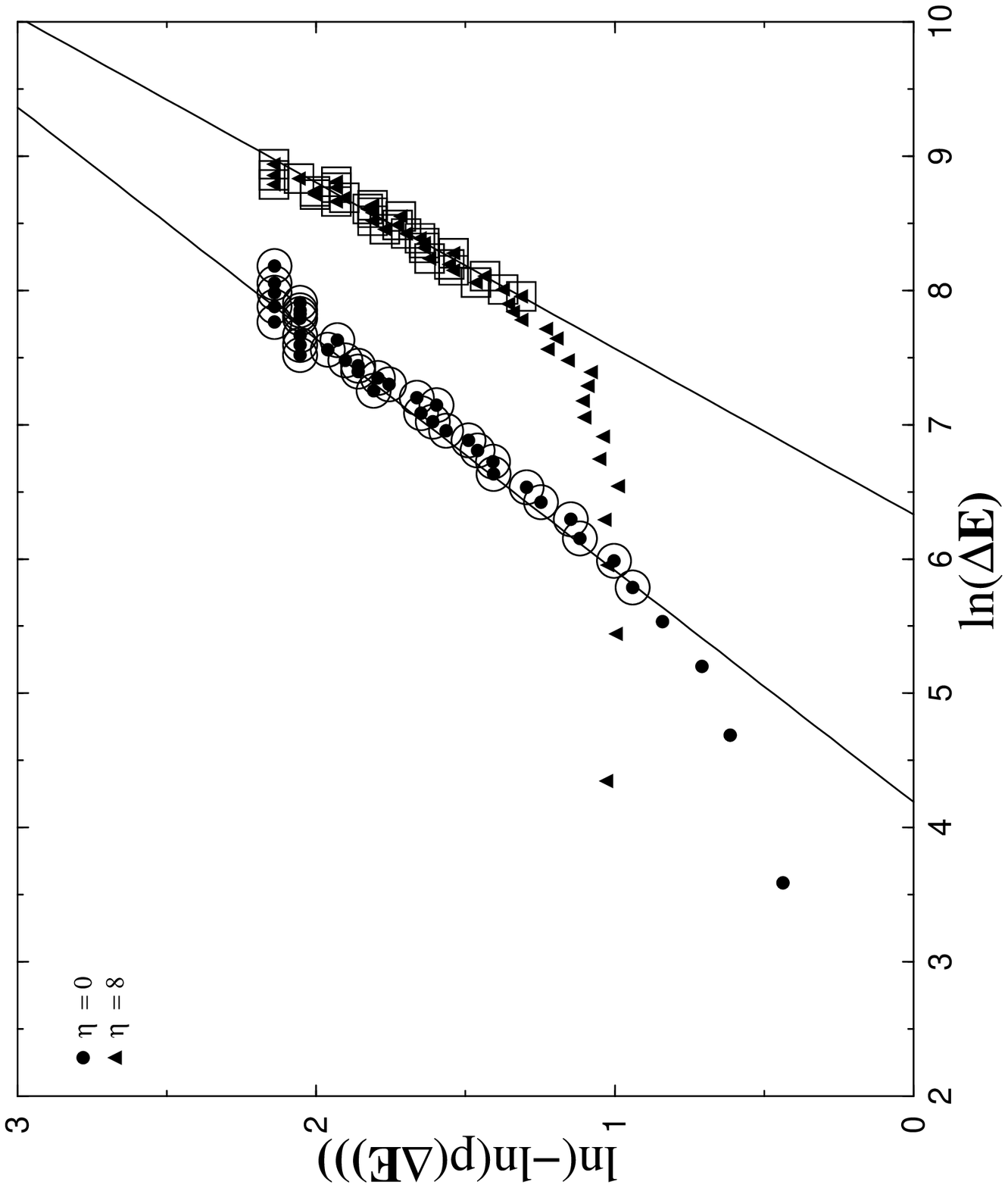,width=15cm,height=20cm,angle=0}
\end{center}
\end{figure}
\bc
{\bf Figure 6b}
\ec

\newpage

\begin{figure}
\begin{center}
\leavevmode
\psfig{figure=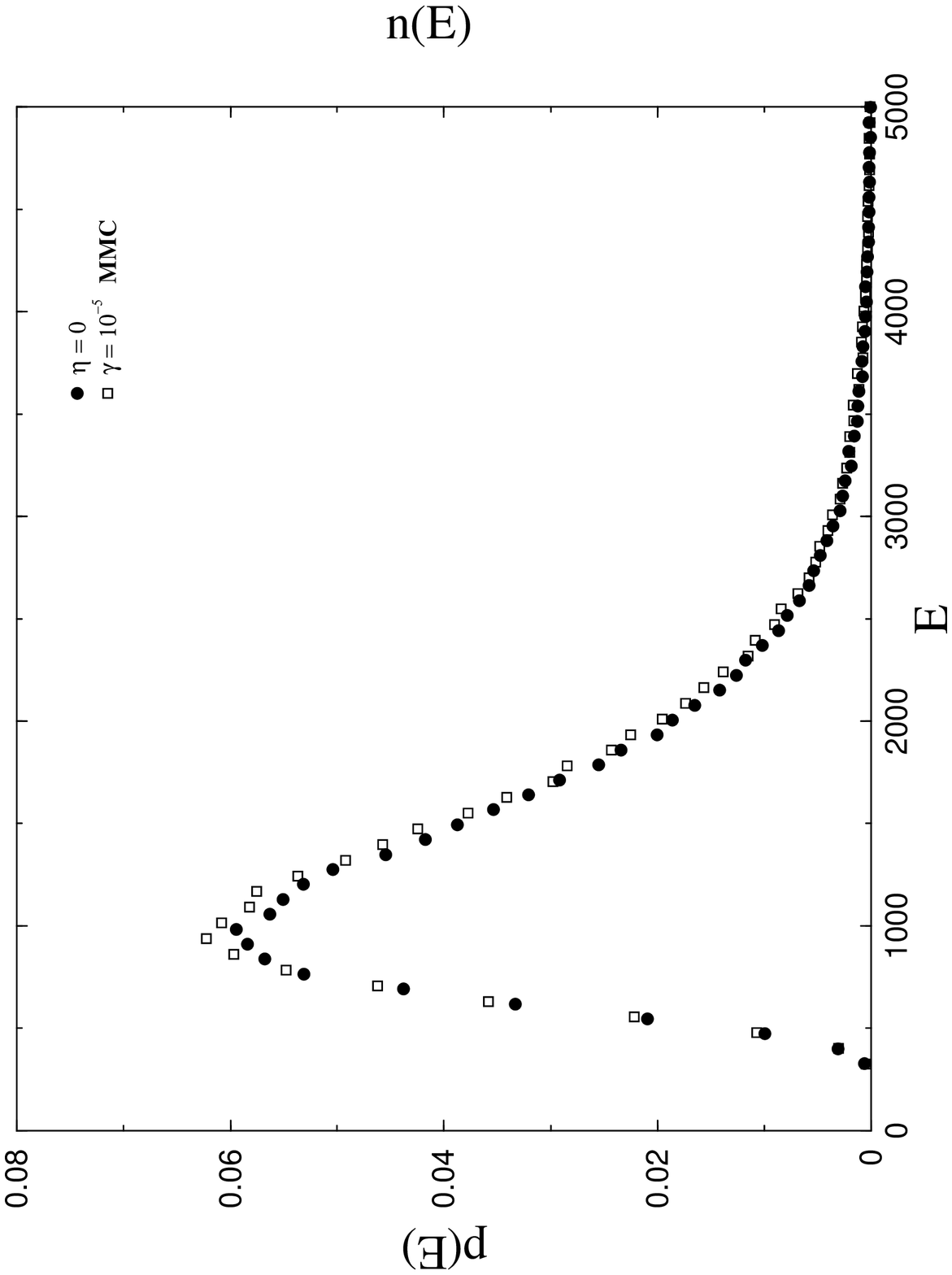,width=15cm,height=20cm,angle=0}
\end{center}
\end{figure}
\bc
{\bf Figure 7}
\ec

\end{document}